# A Compact Design of Four-degree-of-freedom Transmission Electron Microscope Holder for Quasi-Four-Dimensional Characterization

*Yizhi Zhang[1,2], Yeqiang Bu[1], Xiaoyang Fang[1], Hongtao Wang[1,2*]*

[1] Center for X-Mechanics, Zhejiang University, Hangzhou 310027, China

[2] Institute of Applied Mechanics, Zhejiang University, Hangzhou 310027, China

[3] Center for High Pressure Science, State Key Laboratory of Metastable Materials Science and Technology, Yanshan University, Qinhuangdao 066004, China

**ABSTRACT**

Electron tomography (ET) has been demonstrated to be a powerful tool in addressing challenging problems, such as understanding 3D interactions among various microstructures. Advancing ET to broader applications requires novel instrumentation design to break the bottlenecks both in theory and in practice. In this work, we built a compact four-degree-of-freedom (three-directional positionings plus self-rotation) nano-manipulator dedicated to ET applications, which is called X-Nano transmission electron microscope (TEM) holder. All the movements of the four degrees of freedom are precisely driven by built-in piezoelectric actuators, minimizing the artefacts due to the vibration and drifting of the TEM stage. Full 360º rotation is realized with an accuracy of 0.05º in the whole range, which solves the missing wedge problem. Meanwhile, the specimen can move to the rotation axis with an integrated 3D nano-manipulator, greatly reducing the effort in tracking sample locations during tilting. Meanwhile, *in-situ* stimulation function can be seamlessly

---
[*] To whom correspondence should be addressed. E-mail address: anmin@ysu.edu.cn (Anmin Nie) and htw@zju.edu.cn (Hongtao Wang)

integrated into the X-Nano TEM holder so that dynamic information can be uncovered. We expect that more delicate researches, such as those about 3D microstructural evolution, can be carried out extensively by means of this holder in the near future.

**1 Introduction**

In 1959, Richard Feynman brought out the question whether the positions of individual atoms in materials could be located by transmission electron microscopy (TEM) [1, 2]. After nearly half a century, three-dimensional (3D) imaging by TEM, *i.e.* electron tomography (ET), has become more and more compelling in multidisciplinary fields, such as biology, chemistry, physics and materials science [3-5]. Very recently, ET in atomic resolution has been realized for selected nano-crystals with great efforts [6-9], which partly answers the question of Feynman. For the advanced ET, the single-axis tilt scheme is frequently used, which benefits from the simple data acquisition and reconstruction methods [3, 10-12]. The ET resolution is related to not only the acquisition quality of imaging data, but also the tilting angle accuracy and range [13]. The 3D reconstruction resolution in the tilting axis direction is no better than that of the original data source. In the perpendicular direction, the resolution also depends on the sampled volume and the number of projections evenly distributed in 180º range [14]. However, the limited tilting range, as constrained by the instrumentation, causes a missing wedge of information, and consequently leads to resolution reduction [13]. Great efforts have been made to tackle the missing wedge problem, especially from the instrumentation side. Two types of commercialized TEM holders have been made available for ET purpose. Regular TEM holders are limited in tilting range because the sample loading of the holder would collide with the pole piece of the TEM in high tilt angle. To solve this problem, the ET-specialized TEM holders narrow down the front end for sample loading to a few millimeters to fit in the limited pole piece gap at high tilt angle, such as Gantan's Model 698, FISCHIONE's Model 2020 and FEI's Single-Tilt Tomography Holders. These holders can rotate the specimen to a high tilting angle usually around $\pm 70°$. The goniometer of regular TEMs does not allow higher tilting angle upto $\pm 90°$, resulting in the missing-wedge problem. Another type of ET-specialized TEM holders have self-driving tilting function inside the holder. For example, FISCHIONE's Model 2050 holder adopts a 3-position, precision indexing mechanism, which provides the means to orient the specimen in 120º increments. At each increment, the microscope's goniometer is tilted to +60º to acquire a tomographic tilt series. Another example of this type of holder is Mel-Build's HATA-Holder, which can tilt the specimen in 15° increments. In practice, acquiring a few tilt series by manually large-angle rotation makes extra complication in the post-processing alignment and reconstruction. Beside the missing wedge problem, the inaccurate angle readout leads to image artefacts in 3D reconstruction, which become prominent with increasing distance away from the tilt axis [15, 16]. This is a hard problem since there are no easier ways to get accurate angle except the nominal display from the TEM control monitor. A careful calibration reveals that the discrepancy can be as large as $4.0^o$ over an angular range of $\pm\ 90^o$ for typical TEM goniometers [16].

Even with the most advanced aberration-corrected TEM, atomic resolution ET is still a challenging work for the ET community [6, 7]. To push ET to the atomic resolution, one has to re-examine the experimental errors such as vibration and drift caused by tilting the whole TEM stage. It is necessary to re-consider the holder design so that the ET can be a more powerful tool for unveiling the unprecedented information of microstructures. In this work, we built a compact four-degree-of-freedom (positioning in X, Y, Z-directions plus self-rotation) nano-manipulator dedicated to ET applications, named as X-Nano. Full $360^o$ rotation has been realized with an accuracy of $0.05^o$ in the whole range, which solves the missing wedge problem. Meanwhile, the specimen can be placed to the rotation axis with an integrated 3D nano-manipulator, greatly reducing the effort in tracking sample locations during tilting. All the movements of the four degrees of freedom are precisely controlled by piezoelectric actuators, minimizing the artefacts due to vibration and drifting of the TEM stage. We hope, by further improving the precision and stability of the X-Nano holder, more delicate researches, such as those about atomic ET, can be carried out extensively in the future.

**2 Design**

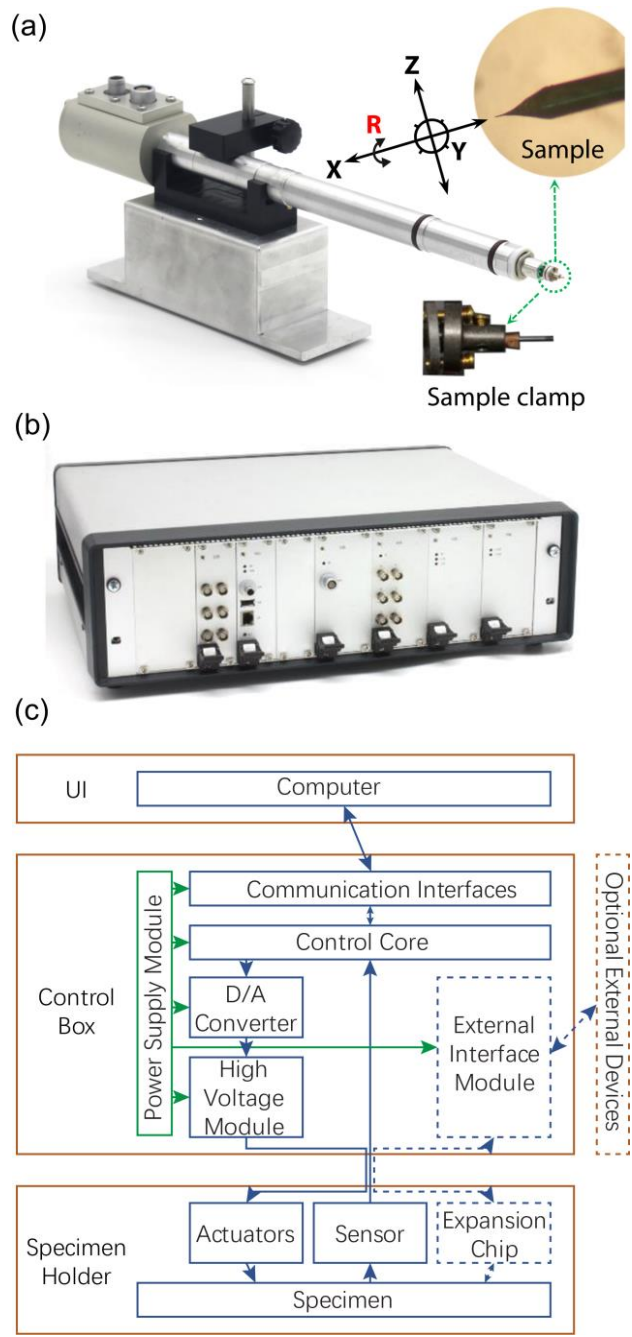

**Figure 1** (a) The X-Nano holder and (b) the control box. Insets to (a) are the sample clamp, a needle-like sample, and the actuator for four degrees of freedom. (c) The X-Nano system layout consists of a host computer, a control box and an X-Nano holder. Each rectangular box represents an independent module. The dashed box is reserved for future extensions for *in-situ* TEM applications.

Figure 1 shows the setup of four-degree-of-freedom nano-manipulator, named as X-Nano, for use inside a JEOL TEM with either ultrahigh resolution or high-resolution pole piece. Full 360°-rotation with angle resolution better than 0.05° is realized with compact piezoelectric actuation. The sample for tomography is preferred to have a needle shape so that it can meet the condition for electron transparency all the time during 360° rotation around the X-axis (upper inset to Fig. 1 (a)). The sample clamp (lower inset to Fig. 1(a)) is specially designed to be fully compatible with the atom probe specimen so that the sample can be used for both TEM and atom probe tomography with no extra effort on modification. In this way, three-dimension structural and elemental characterization of the same sample can be overlaid, providing the comprehensive information. For the ET application, the most tedious work is to track the sample location for the large number of rotation angles, especially when the sample is located far from the geometric rotation axis. Due to the large inertia of the TEM goniometer, the specimen needs a long time to settle down for high resolution imaging. In the present design, the sample can move in all three directions with sub-nanometer accuracy and millimeter motion range using the integrated piezoelectric actuators. Even after a fast and large motion, the sample stops instantly once the electrical driving voltage is held. Figures 1(b) and 1(c) show the control box and the corresponding modular design layout, respectively. Each solid-line box represents an independent module realizing one specific function. The dashed box is reserved for future extension for *in-situ* TEM applications. The arrows indicate the flow direction of signals. The communication between the control box and the host computer obeys the standard User Datagram Protocol (UDP) through a network cable. The user interface is provided in Fig. S1 with full control over high precision movement mode or large range movement mode. Six-channel amplified voltage signals are sent to the piezoelectric actuators independently to realize motions in all four degrees of freedom. The angle readout is sampled at a frequency of 300 Hz in real time *via* Serial Peripheral Interface (SPI) protocol.

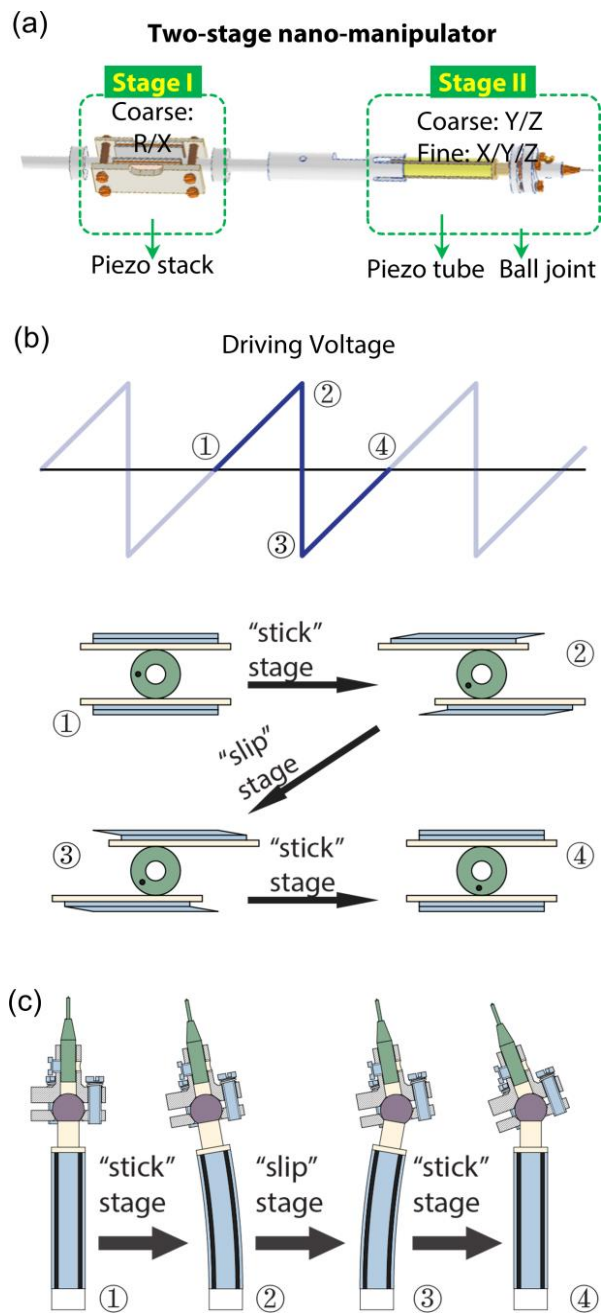

**Figure 2** (a) The two-stage nano-manipulator is designed to combine R/X and Y/Z actuators in series. The fine movement in X/Y/Z direction is realized solely by the piezoelectric tube scanner. (b) A scheme to illustrate the inertial sliding mechanism for rotation. The actuation process consists of four states in one period, numbered as 1 to 4 in sequence. The corresponding positions are marked in the periodical driving voltage profile. (c) A scheme to illustrate the inertial sliding mechanism in the realization of displacement along Y- or Z-direction.

The X-Nano holder move the sample in four degrees of freedom with a two-stage compact nano-manipulator based on the inertial slider mechanism, as shown in Fig. 2(a). The actuator in stage I consists of a stack of two piezoelectric plates polarized in orthogonal in-plane directions, which actuates full 360° rotation (denoted as R) and translational motion along the rotation axis, *i.e.* X-direction. In general, one "stick-slip" period has four sequential states. Figure 2(b) sketches the rotation process for one period. Starting from state 1, the driving voltage changes slowly and the piezoelectric plates on both sides of the central shaft gradually shear, which rotates the shaft for a finite angle ($\Delta\theta$) until state 2. Then the voltage reverses and the plates rapidly shear to the opposite direction, as in state 3. Finally, the voltage gradually changes back to zero again at state 4, and the next driving period starts. Here a well-tuned pressure is applied between the shaft and the piezoelectric plates so that the shaft "sticks" to the plates on slow shear from state 1 to state 2, and relative "slip" on rapid shear from state 2 to state 3. If the slew rate of the driving voltage is high enough, the reverse "slip" rotation can be neglected compared to $\Delta\theta$. In our device, the slew rate achieves about 60 V/μs. We note that $\Delta\theta$ is generally in the order of $10^{-3}$ degree. Repeating the "stick-slip" step achieves stable and continuous rotation, which is critical for ET application. Driving the movement in X-direction is similar to rotation, but it is achieved with two others piezoelectric plates shearing along the

X-direction. The large range translational movement on the horizontal and vertical radial directions, *i.e.* the Y/Z-direction, is realized by another actuator in stage II, which has a piezoelectric scanner tube attached with a ceramic ball. Figure 2(c) schematically shows the corresponding four states in the sticking-slip process of the actuator. The movements in Y and Z directions are slightly coupled with each other due to gravity. The coupling can be minimized by increasing the friction force between the sliding pair with a relatively large pressure. As compared to the commercialized Nanofactory® STM-TEM holder[17], the X-direction displacement is fully decoupled from Y/Z-direction movements in our device.

## 3. Nano-manipulator operation

The stableness of the mechanical system is critical for high resolution TEM imaging. Figure 3 shows a series of bright field TEM images of a $SnS_2$ nanoparticle at magnification from 8k to 600k in a JEOL 2100 TEM equipped with a Gatan 831 CCD camera. The exposure time is set to 1 s. No blur has been noticed at different magnification. The lattice fringe can be clearly observed at both 500k X and 600k X, demonstrating excellent stableness of the two-stage nano-manipulator.

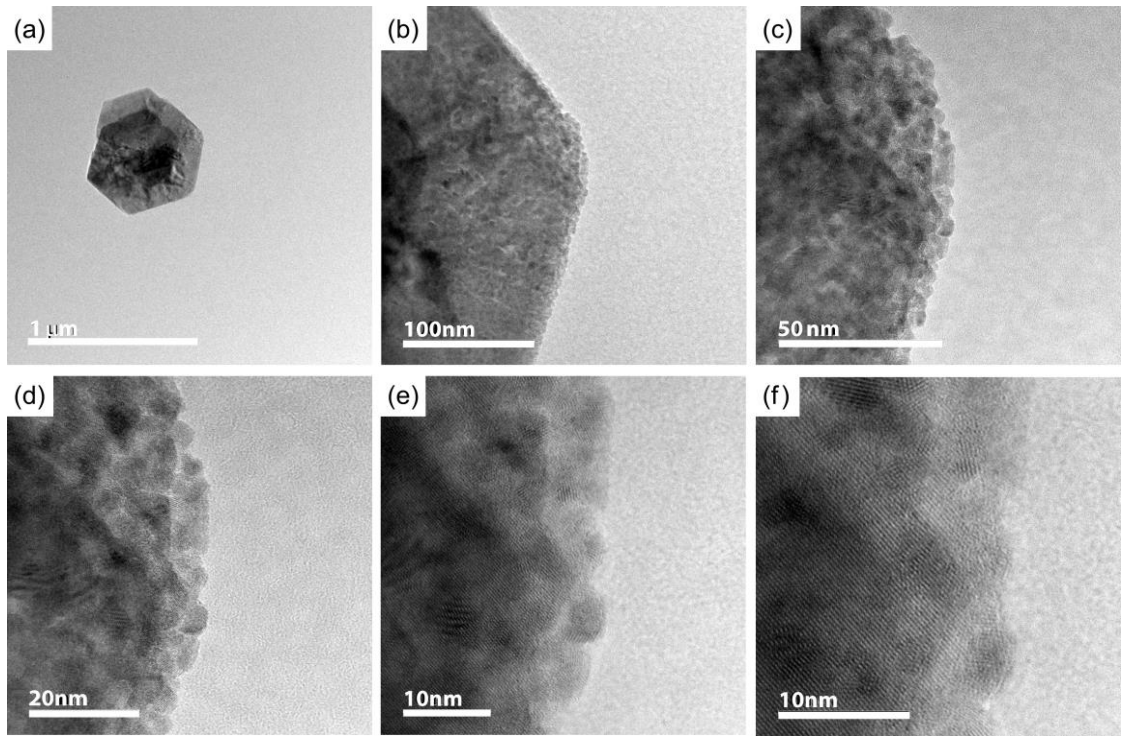

**Figure 3** The bright field TEM images of a $SnS_2$ nanoparticle are captured by using the X-Nano holder at magnifications of (a) 8k X, (b) 60k X, (c) 150k X, (d) 250k X, (e) 400k X, and (f) 600k X.

The large range movement along the X-direction is tested using a needle-like sample in TEM mode at a magnification of 100k X. The tip apex with a dark contrast is taken as a marker, as circled in Fig. 4, to track the position during actuation. Backward and forward motions along X-direction are recorded and supplied as supporting materials (Movie S1). A series of four successive video clips are shown in Fig. 4(a-d). The four locations of the triangle tip marked with circles line up. When the movement is reversed, the tip follows approximately the same track as indicated by the line. Repeating this operation, as demonstrated in Movie S1, one hardly finds any obvious change in the track, suggesting excellent control over the movement in X-direction. Since the R/X and Y/Z movements are decoupled in the design, the image focus of the tip does not change during a repetitive 30 s X-direction motion.

The slight blur is observed only during the actuation. The image becomes clear once the tip stops moving, showing good stableness. No inertial effect is noticed in the operation of the nano-manipulator, which is known as a notorious problem for TEM goniometers. Moving with a large step size is also tested in a low magnification TEM mode (Movie S2). Fast and continuous motion can be achieved in a large range. The magnitude of each step is decided by the driving voltage amplitude (Fig. 4(e-f)), given the saw-tooth driving voltage source. The step size is estimated by averaging displacement of the tip apex for 100 voltage periods. For the large range movement in X-direction, the step size is 5 nm at a driving voltage of 70 V. Below this voltage, the sample makes no accumulated displacement. The step size increases monotonically from 5 nm to 80 nm with voltage ranging from 70 V to 120 V (Fig. 4(e)). A similar trend is observed in driving the Y/Z-direction movements (Fig. 4(f)). However, a lower threshold voltage (30 V) and a much larger step size are observed in the movements of Y/Z-directions. The minimum and maximum step sizes are 0.5 μm and 7 μm, respectively. This is mainly due to the fact that the piezoelectric tube scanner moves much more in the bending mode (± 7 μm at 120 V) than the piezoelectric plate in the shearing mode (± 200 nm at 120 V).

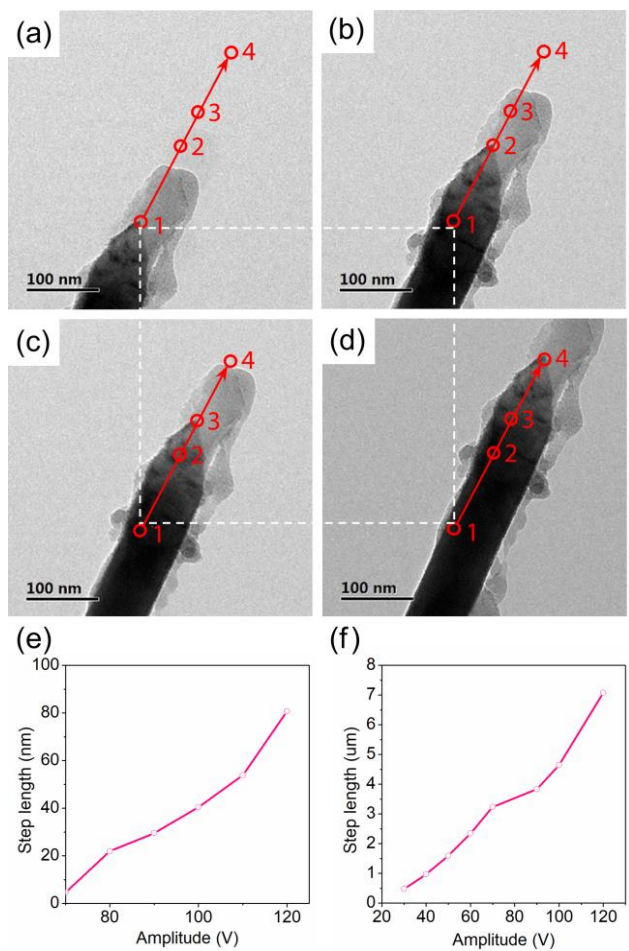

**Figure 4** (a-d) A series of four successive video-clips show the large range movement along X-direction. The video is provided as supplementary material Movie S1. The dashed square shows no relative shift of the markers. (e) The step length of the large range movement along X-direction versus the driving voltage amplitude. (f) The step length of the large range movement along Y-direction versus the driving voltage amplitude.

Due to the capability of three-dimensional positioning, the sample tip displacement can be accurately compensated in rotating operation without using goniometer. However, this exercise can be minimized if the sample is pre-aligned to the rotation axis according to the procedure in Fig. 5(a). We assume that the readouts of X/Y/Z positions in the TEM control panel are all zero after inserting the X-Nano TEM sample holder. Before alignment, the sample tip is brought to the center of the

TEM view screen using the nano-manipulator with large range movement in X- or Y-direction. Assisted by the TEM image focus, the tip height can be properly adjusted by moving in the Z-direction using the nano-manipulator. In general, the rotation axis may not coincide with the sample tip. A three-step alignment procedure is designed to bring the sample tip to the rotation axis. For clarification, we set a global coordinate system with the geometric rotation axis and the electron beam direction being the x- and z-axes, respectively. The first step is to rotate the sample for 180º. The rotation axis must coincide with the section line in the projected view screen (x-y plane of the global coordinate) between 0º and 180º position of the sample, numbered as 1 and 2, respectively. The dots on the left upper and lower dashed circles indicate the corresponding positions in the y-z plane. The second step is to move the tip to the section line. In the y-z plane, the trace of the tip is indicated by the arrow in the right upper circle. After this step, the tip is aligned to the rotation axis in the z-direction. The last step is to rotate the sample for 90º in the reverse direction. The deviation gives the distance between the tip and the rotation axis. Ideally, moving the sample to the section line will make it coincide with the rotation axis, as shown in right lower circle. In practice, the three-step alignment procedure needs to be repeated in TEM mode from low magnification, *e.g.* 200 X, to high magnification, *e.g.* 40k X. Since the minimum step size is about 500 nm in Y-direction for the large range movement, no further improvement can be achieved by aligning the tip to the rotation axis at even higher magnification. In our practice, the alignment process takes about 30 min. Better alignment can be expected if the fine movement is applied. The automation of the procedure is currently under development.

Figure 5 (b-d) show the video clip of a sample tip at 0°, 45° and 90°, respectively. The 90° rotation process is provided as Movie S3. During rotation, the displacement in Y/Z-direction is not compensated in order to measure the accuracy of the alignment procedure. Two dashed lines are plotted in parallel to the rotation axis and through the tip apex at 0° and 90°, respectively. The transverse displacement is measured to be 300 nm, which is comparable to the minimum step size in Y-direction for the large range movement. No obvious focus change is observed in the rotation at the given magnification. The stability of the eucentric position of the specimen during its rotation directly influences the spatial resolution of the resulting tomography. The mechanical drifting is measured by using a nano needle. After alignment, the projection of the sample tip deviation from the rotation axis is measured in a full 360° rotation without any position compensation (Fig. S3). The maximum eucentric position change is around 500 nm, which can be compensated by the fine movement of the nanomanipulator. Beside the allowed range of rotation, the angle accuracy and the rotation step size are two important factors for TEM tomography application. The angular sensor is designed on the basis of Hall effect. It is calibrated by the industrial photoelectrical encoder with an angle accuracy of 0.02°. The accuracy is determined to be 0.05° in full 360° range (Fig. S2). The step size is measured by averaging over 400 voltage pulses (Fig. 5(e)). The threshold voltage is 50 V and the corresponding minimum magnitude is $8 \times 10^{-6}$ °. The step size monotonically increases with the voltage amplitude and reaches $4 \times 10^{-3}$ ° at 120 V. Since the high-speed SPI protocol enables fast angle readout sampling at a frequency of 300 Hz, a PID control is easily coded in the software for automatic rotation to the set point.

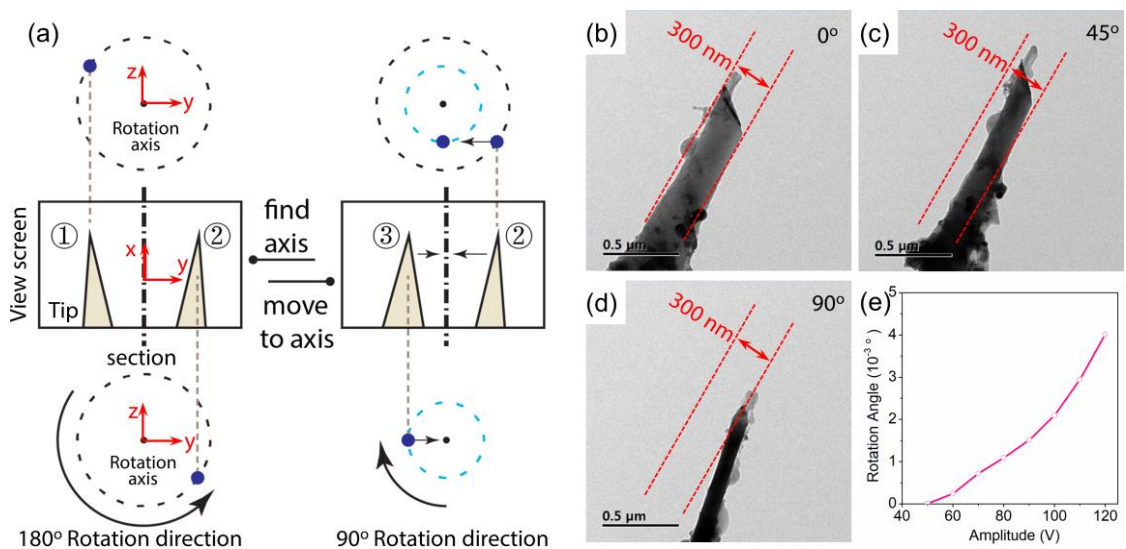

**Figure 5** (a) A schematic illustrates the alignment procedure. The sample position is shown in both x-y plane (the view screen) and y-z plane. A series of three video clips show the TEM images at tilting angles of (b) 0º, (c) 45º and (d) 90º. The video is provided as supplementary material Movie S3. (e) The step length of rotation depends on the driving voltage.

## 4 Applications of X-Nano TEM holder

Applications of the X-Nano TEM holder have been proposed by research groups in the fields of material science. In a recent publication, Tian *et al*. at Yanshan University have achieved the elastic tensile strain of 13.4% in the <100>-oriented diamond nanoneedles using the nano manipulation capability of this setup [18]. These values are comparable with the theoretical elasticity and Griffith strength limits of diamond. Gao *et al*. at Zhejiang University have performed *in-situ* manipulation and Joule welding of graphene assemblies [19]. However, integration nano manipulation with 360º rotation around the tilt axis enables possible exploration in uncover the change in 3D microstructure upon *in-situ* stimulus, which is still a challenging task. It is noted that a pioneer work by Kacher and Robertson has shown the quasi-four-dimensional analysis of dislocation interactions with grain boundaries in 304 stainless steel [20]. The 3D reconstruction of dislocation arrays was especially highlighted in their work for its unparalleled role in uncovering the mechanism of dislocation interactions. However, the tomography was only used as a post-mortem analysis technique, which required sample transfer between *in-situ* and tomography holders. As commented in a review on the electron tomography of dislocation structures from the same research group, the coupling of tomography and *in-situ* TEM deformation was not possible at the time of their study [21]. Since there was no opportunity to see how the microstructure evolved with further loading, some in-depth insight could not be verified by carrying post-deformation tomography [21]. Exploring in this direction, we herein propose a new method coupling both *in-situ* TEM deformation and tomography with the X-Nano holder.

Figure 6 shows an array of dislocations produced in silicon nanopillar by nano-compression in TEM. It is noted that only a small strain is applied to the sample so that only a few dislocations are injected from the contact. In this way, the crystal orientation remains nearly unchanged, which guarantees the successful characterization of dislocations. As the dislocation slip in a planar form, we propose a procedure simplifying the 3D reconstruction by utilizing the accurate angle control of the X-Nano holder ( $\pm 0.05º$ in full 360º range), as following:

Step 1. Collect the tilting series of the dislocations. Choose one set of dislocations and determine the tilting angle α when the dislocation is in the edge-on direction. In this way, the slip plane can be determined.

Step 2. Rotate the sample to the angle of α + 90º, at which the electron beam is perpendicular to the slip plane. The locations of the dislocation lines are measured. The 3D configurations of the dislocation lines are thus obtained by coordinate transformation.

Step 3. Verify the 3D reconstruction by overlap its projection with TEM images recorded at various angles.

The dislocation array was labelled as A after the initial loading. The sample was rotated from 0º to 180º at 1º increment around the [-111] axis, which shown the spatial configuration of dislocation array A. Figure 6 (a) shows the selected dark-field TEM images at rotating angles of 180º, 140º, 110º and 50º, respectively. The 3D arrangement of the dislocation array can be reconstructed according to the proposed procedure, as shown in Fig. 6 (b). To show the consistency, the projection of the reconstructed 3D dislocation arrays is overlapped with the original dark field TEM images in Fig. 6 (a). More details can be found in the supplementary Movie S4. The dislocation array A was rotated at the edge-on state at the angle of 50º, and the slip plane was determined to be (111) plane combined with the corresponding selected area diffraction (SAED) pattern. With further compression, there is another dislocation source activated on the fracture surface, and produced dislocation array B as shown in Fig. 6 (c). The corresponding dislocation arrangement is shown in Fig. 6 (d) and Movie S5. The sample which received twice loadings was rotated again, and the spatial distribution of dislocation array A and B was displayed. The slip plane of dislocation array B was determined to be (1-11) plane according to the SAED at 110º. Here, the functions of *in situ* loading and three-dimensional rotating of X-Nano TEM stage help us to realize the three-dimensional evolution of dislocation nets. As the sample is pre-aligned to the rotation axis, the effort for tracking position is greatly reduced during the large-angle rotation. By using the X-Nano TEM holder, the 3D tomography may become a routine task.

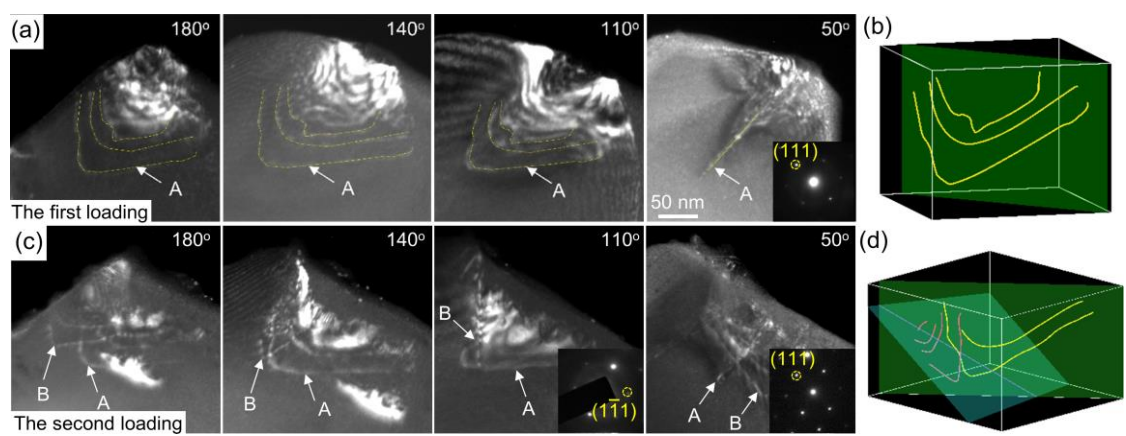

**Figure 6** The operations for four-dimensional TEM. Dislocation images of the samples (a) after the first loading and (c) after the second loading at the rotating angle of 180º, 140º, 110º and 50º, respectively. Insets are corresponding selected electron diffraction patterns. The reconstructed dislocation arrays are shown in (b) and (d), respectively. To show the consistency, the projection of the reconstructed 3D dislocation arrays is overlapped with the original dark field TEM images in Fig. 6 (a). More details can be found in the supplementary Movie S4 and S5.

**5 Conclusions and outlook**

Electron tomography has been demonstrated to be a powerful tool in addressing challenging problems such as understanding 3D interactions among various microstructures. Advancing ET to higher resolution and broader applications requires novel instrumentation to break bottlenecks both in theory and in practice. In this work, we have built the dedicated piezoelectric-actuated X-Nano holder for ET applications. Full 360º rotation has been realized with an accuracy of 0.05º in the whole range, which solves the missing wedge problem. Stage vibration and drifting are minimized by using the precise piezoelectric actuators. Meanwhile, the effort is greatly reduced in sample location tracking during tilting after a simple pre-alignment operation. For the current being, we can only realize the sequential characterization of *in-situ* deformation and tomography. The real time observation of 3D microstructure evolution is still far beyond our capability. On the other hand, the theoretical study goes even further, *e.g.* supersonic dislocations gliding [22]. To minimize the gap, it is crucial to revolutionize TEMs, which have been so successful since 1940s.

*This work was supported by the financial support from the Natural Science Foundation of China (Grant No. 11672355 and 11725210).*

1    Hey A. Feynman And Computation. Westview Press, 2002.

2    Xu R, Chen C-C, Wu L*, et al.* Three-dimensional coordinates of individual atoms in materials revealed by electron tomography. Nat Mater, 2015, 14: 1099

3    Peter E, Osama A, J. R M*, et al.* Electron Tomography: A Three-Dimensional Analytic Tool for Hard and Soft Materials Research. Adv Mater, 2015, 27: 5638-5663

4    Weyland M, Midgley P A. Electron tomography. Mater Today, 2004, 7: 32-40

5    Midgley P A, Dunin-Borkowski R E. Electron tomography and holography in materials science. Nat Mater, 2009, 8: 271

6    Miao J, Ercius P, Billinge S J L. Atomic electron tomography: 3D structures without crystals. Science, 2016, 353:

7    Bals S, Goris B, De Backer A*, et al.* Atomic resolution electron tomography. MRS Bull,

2016, 41: 525-530

8   Chen C-C, Zhu C, White E R, *et al.* Three-dimensional imaging of dislocations in a nanoparticle at atomic resolution. Nature, 2013, 496: 74

9   Scott M C, Chen C-C, Mecklenburg M, *et al.* Electron tomography at 2.4-ångström resolution. Nature, 2012, 483: 444

10  Fernández J J, Lawrence A F, Roca J, *et al.* High Performance Computing in Electron Microscope Tomography of Complex Biological Structures. Berlin, Heidelberg: Springer Berlin Heidelberg, 2003.

11  Slater T J A, Janssen A, Camargo P H C, *et al.* STEM-EDX tomography of bimetallic nanoparticles: A methodological investigation. Ultramicroscopy, 2016, 162: 61-73

12  Winkler H, Taylor K A. Marker-free dual-axis tilt series alignment. J Struct Biol, 2013, 182: 117-124

13  Midgley P A, Weyland M. 3D electron microscopy in the physical sciences: the development of Z-contrast and EFTEM tomography. Ultramicroscopy, 2003, 96: 413-431

14  Hart R G. Electron Microscopy of Unstained Biological Material: The Polytropic Montage. Science, 1968, 159: 1464-1467

15  Hayashida M, Malac M, Bergen M, *et al.* Accurate measurement of relative tilt and azimuth angles in electron tomography: A comparison of fiducial marker method with electron diffraction. Rev Sci Instrum, 2014, 85: 083704

16  Hayashida M, Terauchi S, Fujimoto T. Calibration method of tilt and azimuth angles for alignment of TEM tomographic tilt series. Rev Sci Instrum, 2011, 82: 103706

17  Svensson K, Jompol Y, Olin H, *et al.* Compact design of a transmission electron microscope-scanning tunneling microscope holder with three-dimensional coarse motion. Rev Sci Instrum, 2003, 74: 4945-4947

18  Nie A, Bu Y, Li P, *et al.* Approaching diamond's theoretical elasticity and strength limits. Nat Commun, 2019, 10: 5533

19  Liu Y, Liang C, Wei A, *et al.* Solder-free Electrical Joule Welding of Macroscopic Graphene Assemblies. Mater Today Nano, 2018, 3: 1-8

20  Kacher J, Robertson I M. Quasi-four-dimensional analysis of dislocation interactions with grain boundaries in 304 stainless steel. Acta Mater, 2012, 60: 6657-6672

21  Liu G S, House S D, Kacher J, *et al.* Electron tomography of dislocation structures. Mater Charact, 2014, 87: 1-11

22  Peng S, Wei Y, Jin Z, *et al.* Supersonic Screw Dislocations Gliding at the Shear Wave

Speed. Phys Rev Lett, 2019, 122: 045501

# Supporting Materials

**A Compact Design of Four-degree-of-freedom Transmission Electron Microscope Holder for Quasi-Four-Dimensional Characterization**

**Table of Contents**

- FIG. S1   The user interface of the control software in the host computer.
- FIG. S2   The calibration of the angular sensor.
- FIG. S3   The maximum deviation from the rotation axis in 360° rotation
- Movie S1  The small step coarse movement of the sample tip in the X-direction.
- Movie S2  The large step coarse movement of the sample tip in the X-direction.
- Movie S3  The rotation of the sample tip for 90°. The movie is played at 30 X speed.
- Movie S4  The 3D reconstruction of dislocations after the initial in-situ deformation.
- Movie S5  The 3D reconstruction of dislocations after the second in-situ deformation.

Supporting Materials

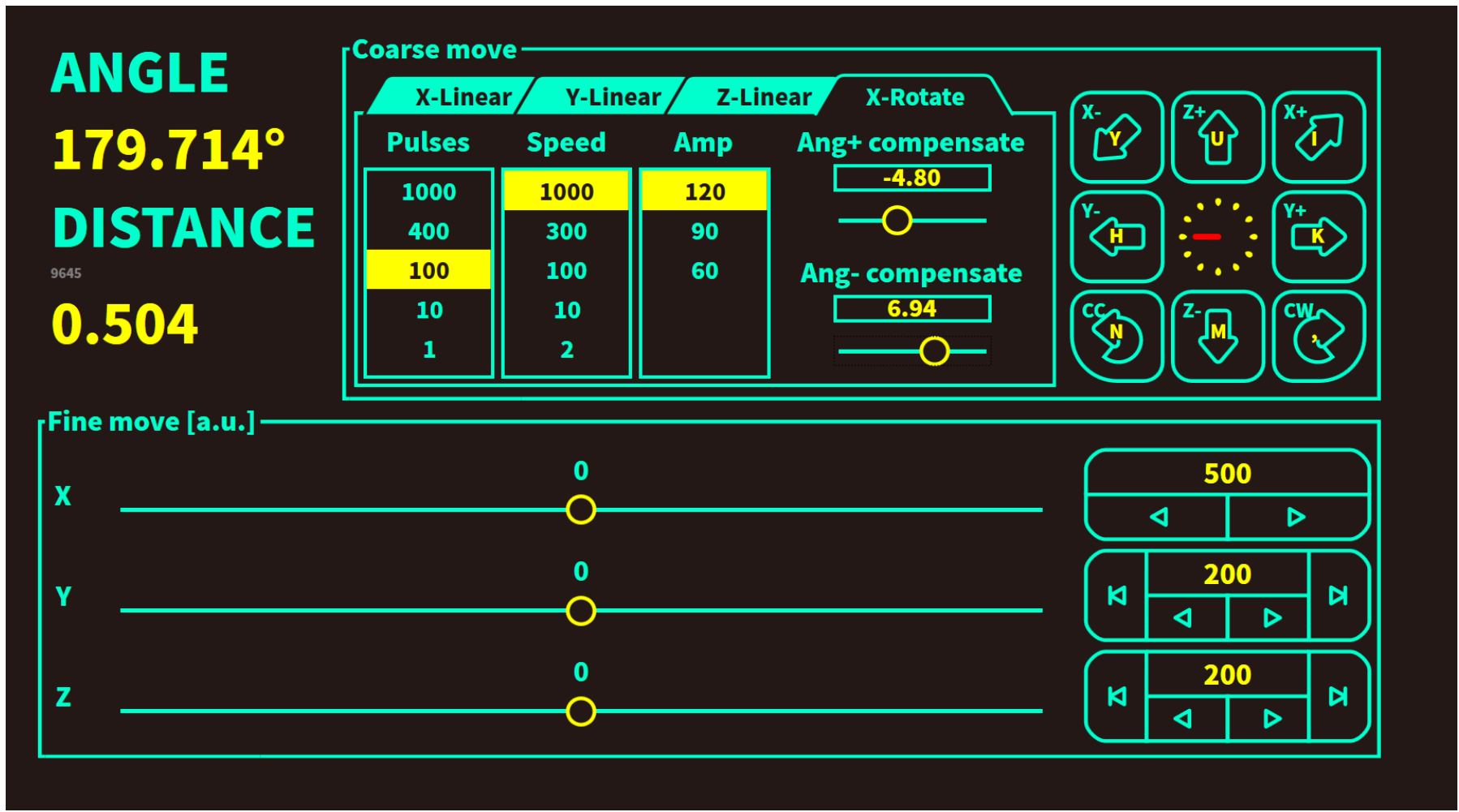

FIG. S1 The user interface of the control software in the host computer.

Supporting Materials

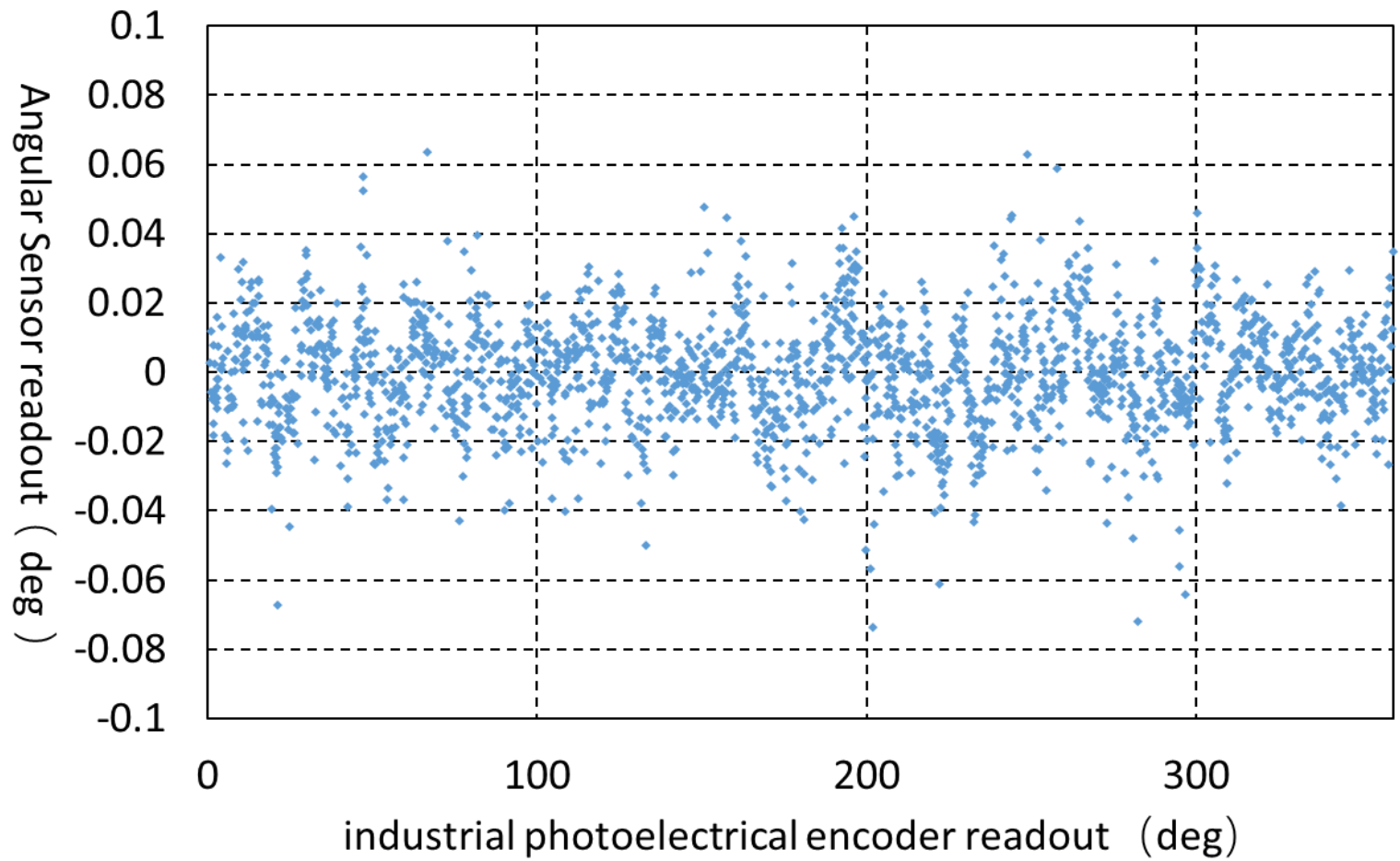

FIG. S2    The calibration of the angular sensor.

Supporting Materials

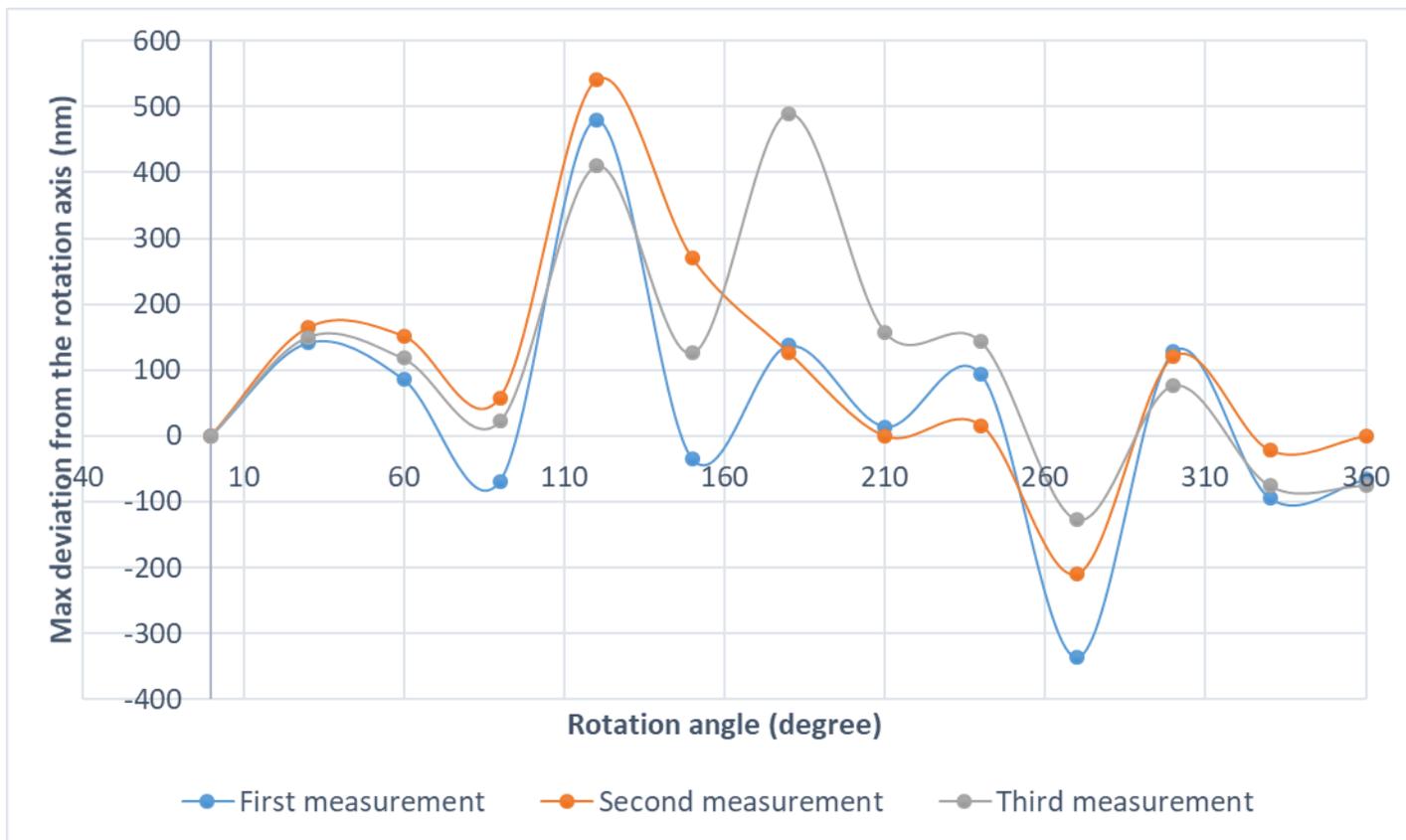

FIG. S3 The maximum deviation from the rotation axis in 360° rotation

Supporting Materials

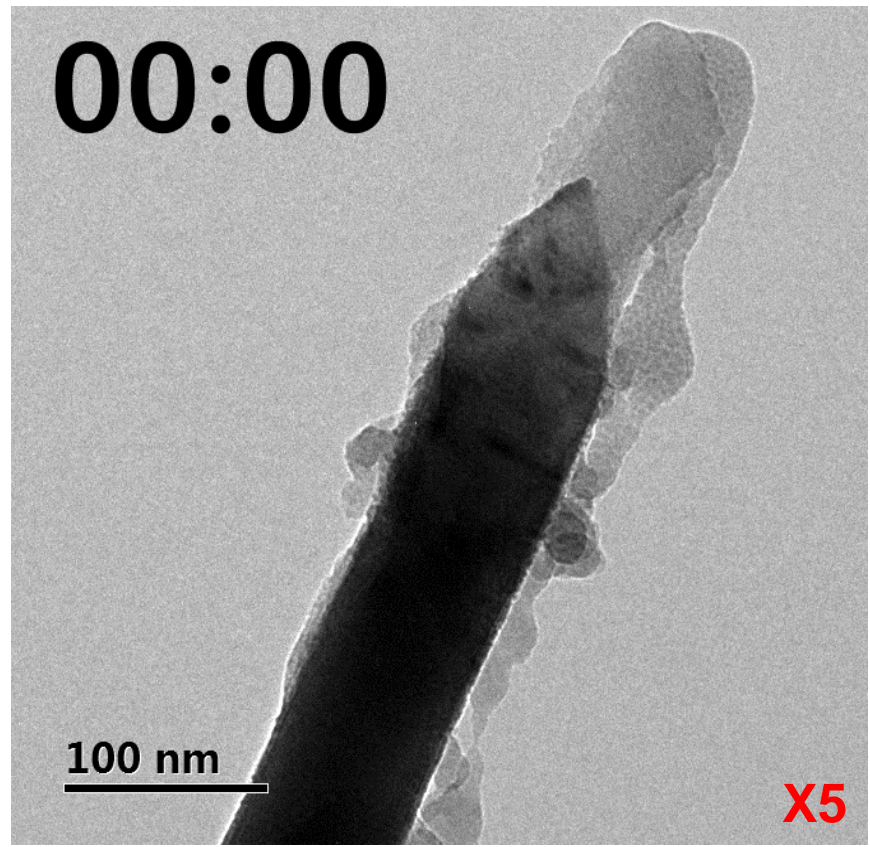

Movie S1  The small step coarse movement of the sample tip in the X-direction.

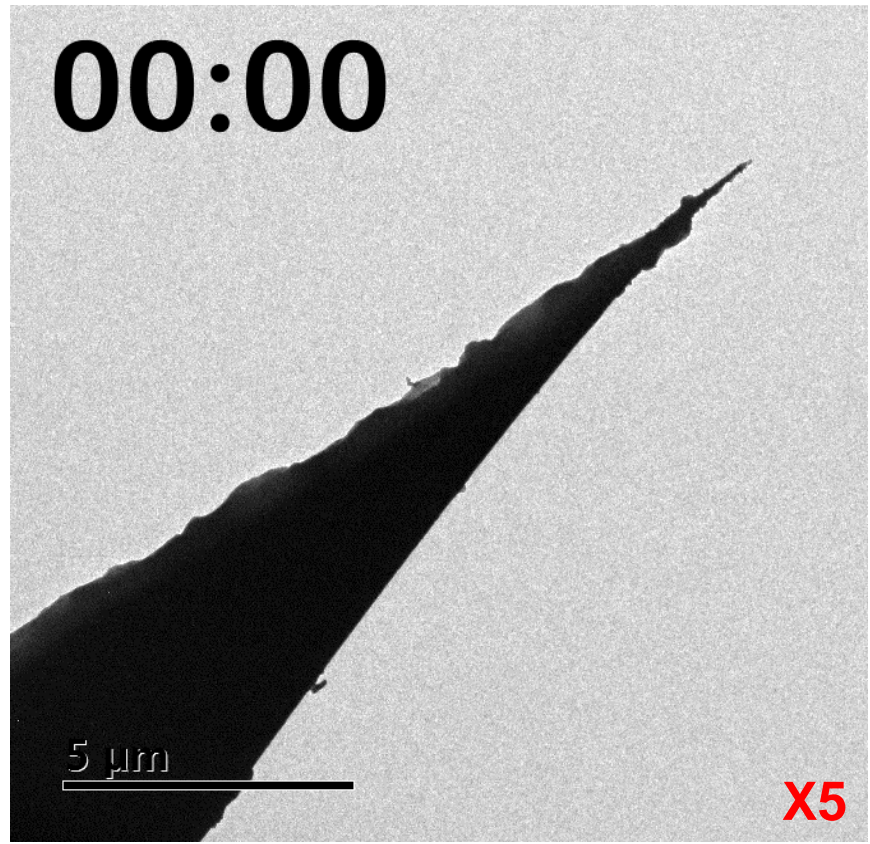

Movie S2  The large step coarse movement of the sample tip in the X-direction.

Supporting Materials

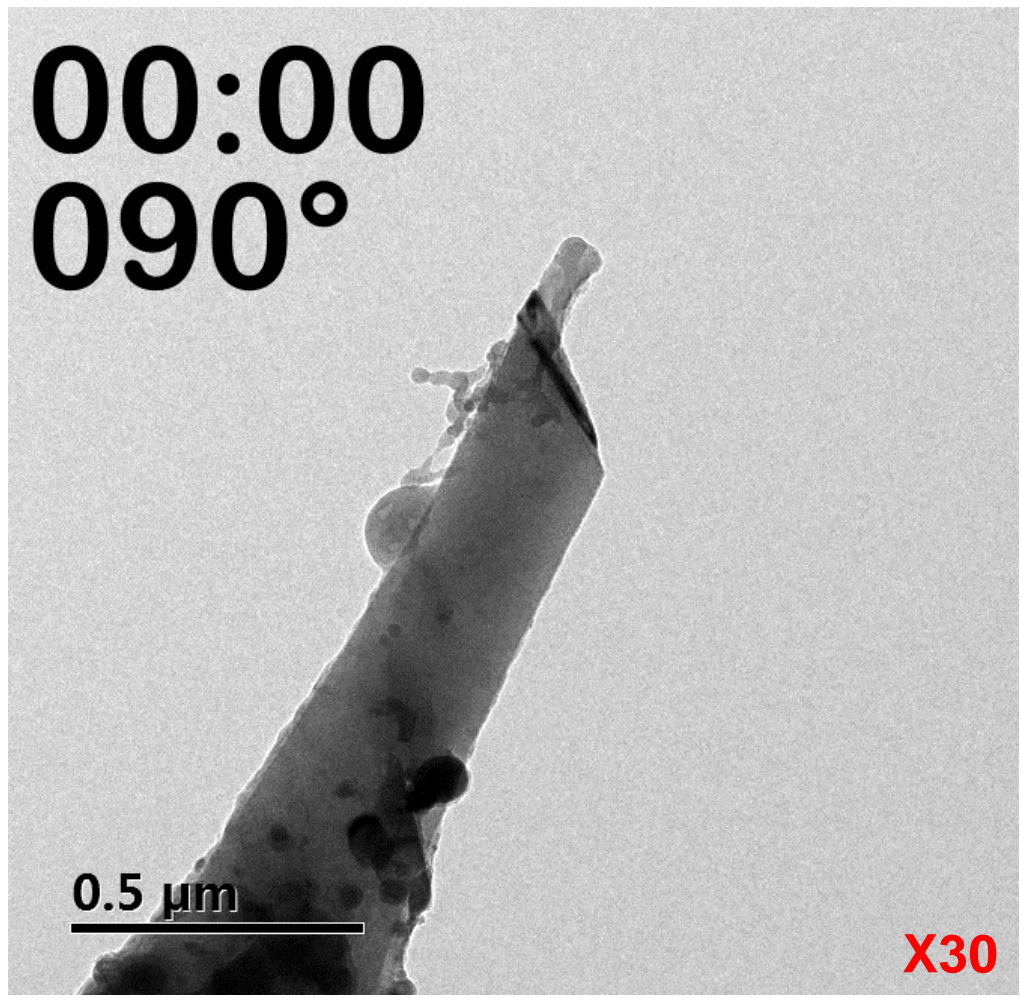

Movie S3  The rotation of the sample tip for 90º. The movie is played at 30 X speed.

Supporting Materials

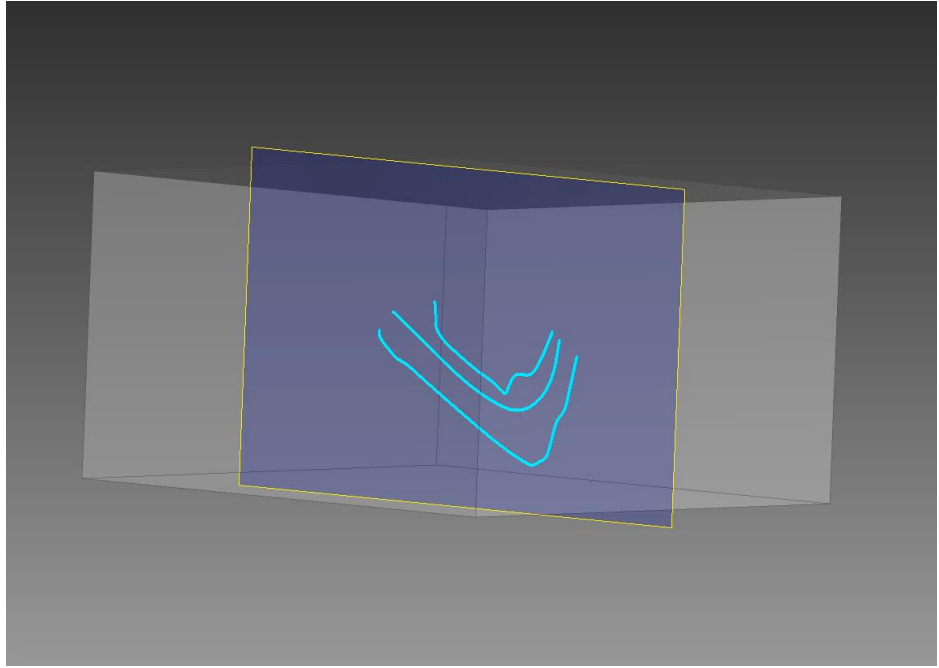

Movie S4  The 3D reconstruction of dislocations after the initial in-situ deformation.

Supporting Materials

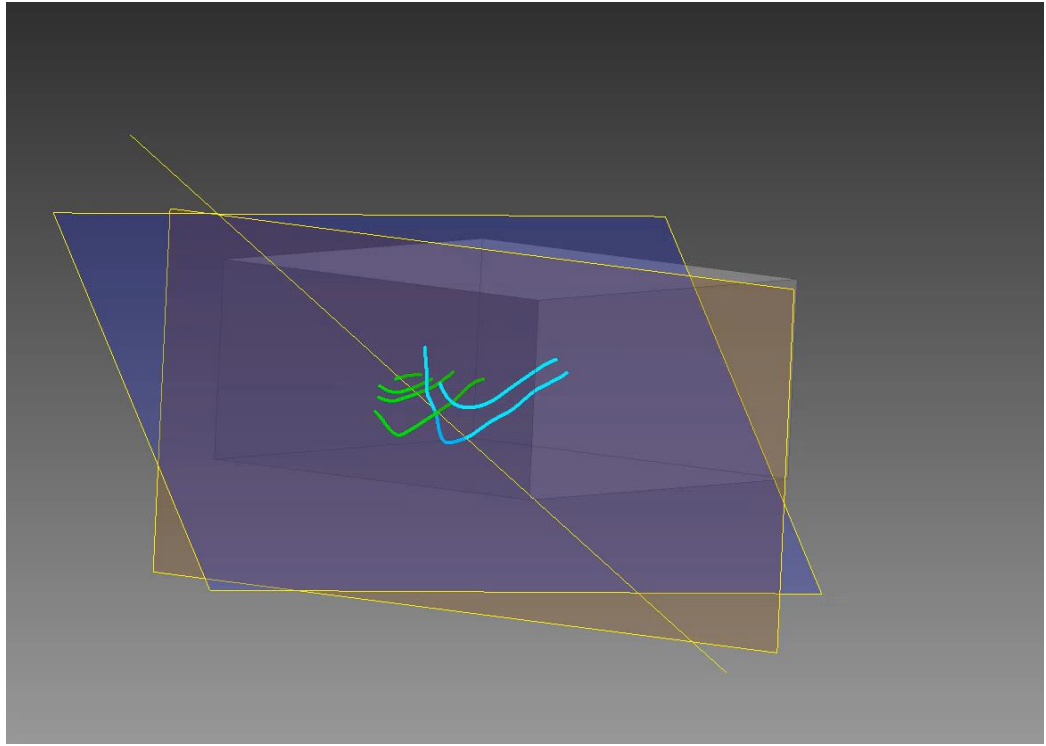

Movie S5  The 3D reconstruction of dislocations after the second in-situ deformation.